\documentclass[10pt,conference]{IEEEtran}
\IEEEoverridecommandlockouts
\usepackage{amsmath,amssymb,amsfonts}
\usepackage{bm}
\usepackage{cite}
\usepackage{algorithmic}
\usepackage{graphicx}
\usepackage{xcolor}
\usepackage{textcomp}
\usepackage{footnote}
\def\BibTeX{{\rm B\kern-.05em{\sc i\kern-.025em b}\kern-.08em
    T\kern-.1667em\lower.7ex\hbox{E}\kern-.125emX}}
\usepackage{booktabs}
\usepackage{tikz}
\usepackage{pgfplotstable}
\usepackage{pgfplots} 
\pgfplotsset{compat=1.9}
\usepackage{hyperref}
\hypersetup{
    colorlinks=true,
    citecolor=blue,
    linkcolor=black,
    urlcolor=black,
    linkbordercolor={1 1 1},
    citebordercolor={1 1 1},
}
\def\x{{\mathbf x}}
\def\X{{\mathbf X}}
\def\Z{{\mathbf Z}}
\def\S{{\mathcal S}}
\def\T{{\mathcal T}}
\newcommand{\overbar}[1]{\mkern 1.5mu\overline{\mkern-1.5mu#1\mkern-1.5mu}\mkern 1.5mu}
\begin{document}

\title{\huge Unsupervised Domain Adaptation for Acoustic Scene Classification Using Band-Wise Statistics Matching \thanks{$^\star$\,A joint institution of the Friedrich-Alexander-University Erlangen-Nurnberg (FAU) and Fraunhofer Institute for Integrated Circuits (IIS).}
}

\author{\IEEEauthorblockN{Alessandro Ilic Mezza\IEEEauthorrefmark{1}, Emanu\"el A. P. Habets\IEEEauthorrefmark{2}, Meinard M\"uller\IEEEauthorrefmark{2} and Augusto Sarti\IEEEauthorrefmark{1}}
\IEEEauthorblockA{\IEEEauthorrefmark{1}Dipartimento di Elettronica, Informazione e Bioingegneria, Politecnico di Milano, 20133 Milan, Italy\\
Email: alessandroilic.mezza@polimi.it, augusto.sarti@polimi.it}
\IEEEauthorblockA{\IEEEauthorrefmark{2}International Audio Laboratories Erlangen$^\star$, Am Wolfsmantel 33, 91058 Erlangen, Germany\\
Email: emanuel.habets@audiolabs-erlangen.de, meinard.mueller@audiolabs-erlangen.de}
}

\maketitle

\begin{abstract}
The performance of machine learning algorithms is known to be negatively affected by possible mismatches between training (source) and test (target) data distributions. 
In fact, this problem emerges whenever an acoustic scene classification system which has been trained on data recorded by a given device is applied to samples acquired under different acoustic conditions or captured by mismatched recording devices.
To address this issue, we propose an unsupervised domain adaptation method that consists of aligning the first- and second-order sample statistics of each frequency band of target-domain acoustic scenes to the ones of the source-domain training dataset.
This model-agnostic approach is devised to adapt audio samples from unseen devices before they are fed to a pre-trained classifier, thus avoiding any further learning phase.
Using the DCASE 2018 Task 1-B development dataset, we show that the proposed method outperforms the state-of-the-art unsupervised methods found in the literature in terms of both source- and target-domain classification accuracy.
\end{abstract}

\begin{IEEEkeywords}
Unsupervised domain adaptation, mismatched recording devices, acoustic scene classification.
\end{IEEEkeywords}

\section{Introduction}
\label{sec:intro}
Acoustic Scene Classification (ASC) is the task of assigning a categorical label to a test audio recording to characterize the environment in which it was captured --- for instance ``Metro station'', ``Park'', ``Airport''.
In recent years, deep learning (DL) has proven to be an essential and powerful tool to effectively tackle this problem \cite{mesaros2018detection, mesaros2018acoustic, sakashita2018, koutini2019receptive}. However, as a downside, DL-based ASC systems tend to be susceptible to the effects of domain shift, i.e., the well-known performance degradation that affects machine learning algorithms when trained and tested on data drawn from different distributions \cite{ben2010theory}. 
Domain adaptation (DA), despite having been extensively investigated in fields such as natural language processing \cite{daume2006domain,blitzer2006domain}, sentiment analysis \cite{blitzer2007biographies,glorot2011domain} and computer vision \cite{saenko2010adapting, hoffman2017cycada}, is still a relatively new topic in the context of ASC.
Since 2018, the IEEE AASP Challenge on Detection and Classification of Acoustic Scenes and Events (DCASE) have included a subtask specifically designed to encourage DA, namely Task 1-B on ``Acoustic Scene Classification with mismatched recording devices.'' In this task each recording device is regarded as a separate domain.
Nonetheless, the training datasets adopted by the DCASE challenges in 2018 \cite{mesaros2018multi} and 2019 \cite{mesaros2019acoustic}, although being highly unbalanced in favour of a single recording device, also contain several acoustic scenes that were simultaneously captured using both source (Device A) and target devices (Devices B and C); we refer to such data as \textit{parallel} data.
As a result, most ASC models found in the literature are trained using both source- and target-domain samples and thus are not \textit{blind} with respect to the target domain prior to the adaptation.

To date, only a few studies (such as \cite{gharib2018unsupervised, drossos2019unsupervised, mun2019domain}) have applied unsupervised DA techniques to ASC models that were trained solely on source-domain data.
In \cite{gharib2018unsupervised} and \cite{drossos2019unsupervised}, the authors propose to adapt a pre-trained DL-based ASC model by means of adversarial learning. In particular, \cite{gharib2018unsupervised} follows the Adversarial Discriminative Domain Adaptation framework presented in \cite{tzeng2017adversarial} to adapt the convolutional layers of a pre-trained CNN so to force the feature extractor into yielding domain-invariant data representations. Furthermore, \cite{drossos2019unsupervised} improves over \cite{gharib2018unsupervised} by replacing the adversarial adaptation process with a module based on Wasserstein Generative Adversarial Networks (WGAN) \cite{arjovsky2017wasserstein}. 
Meanwhile, the authors of \cite{mun2019domain} recently proposed a different paradigm: instead of adapting a pre-trained ASC model, DA is enforced directly on the acoustic scenes using a Factorized Hierarchical Variational AutoEncoder (FHVAE). This method aims to disentangle scene-dependent and channel-related characteristics in terms of a pair of latent variables $z_1$ and $z_2$. Afterwards, a ``channel conversion'' step is performed in the latent space by shifting $z_2$ by a domain-specific factor $\Delta\mu_2$.

The strategy of applying DA on audio data before model training and evaluation is adopted also by the winning submission to DCASE 2019 Task 1-B \cite{komider2019}, where the main idea is to equalize the different frequency responses of mismatched recording devices. To this end, a set of spectral coefficients is computed by dividing the spectra of a matched pair of simultaneous recordings from different devices. Coefficients are then averaged over multiple pairs.
Finally, spectral correction is applied by multiplying each frequency bin of the short-time Fourier transform of every acoustic scene by the coefficient associated to the corresponding frequency band. 

Despite having proven to be quite effective, not only \cite{komider2019} requires several target-domain samples prior to the training phase, but it also makes a further assumption on the availability of parallel audio files. 
In turn, the adversarial DA methods found in the literature \cite{gharib2018unsupervised, drossos2019unsupervised} suffer from two critical limitations. First, they entail a whole new adaptation phase every time a novel target domain is encountered. Second, they require a suitably sized target-domain dataset to train the adaptation module.
The technique presented in \cite{mun2019domain}, while being designed to address the latter shortcomings, requires an additional dataset of acoustic scenes (dubbed ``Universal domain'') in order to pre-train the FHVAE and thus compute the channel conversion parameter $\Delta\mu_2$.
In \cite{mun2019domain}, two additional variants are described: the first one uses source-domain data
for the pre-training, while the second employs target-domain data. 
The latter variant, however, despite being the best performing of the three, violates the requirement of not relying on information from the target devices at training time.
Moreover, the classifier is learnt using the reconstructed features decoded by the FHVAE: the adaptation procedure cannot be readily applied to any previously optimized ASC model, as it inevitably entails a training.

In this paper, we present an effective unsupervised DA procedure for ASC that is capable to overcome the limitations of adversarial strategies by performing DA at data level, but without requiring additional data as in \cite{mun2019domain}.
The main idea is to apply a preprocessing technique prior to the test phase in which the first- and second-order sample statistics of each frequency band of test data are matched to the ones of the source-domain training dataset.
The proposed approach consists of three main steps. First, just before the training phase, the sample mean and standard deviation for each frequency band across every sample in the source-domain training dataset is computed. Second, at inference time, a band-wise standardization is applied to the target-domain test data so to obtain zero-mean and unit-variance frequency bands throughout the dataset. Third, the standardized dataset is finally adapted using the means and standard deviations computed at Step~1.
We show that this procedure can significantly increase the target-domain performance of an ASC system, while having a low computational cost and not substantially affecting the results of source-domain classification.

\section{Proposed Method}
\label{sec:method}

Let $\x\in\mathbb{R}^{M \times K}$ be the spectrogram of an acoustic scene, where $M$ and $K$ represent the number of time frames and frequency bands, respectively. 
Let 
$\X^\S\in\mathbb{R}^{N_\S \times M \times K}$
and
$\X^\T\in\mathbb{R}^{N_\T \times M \times K}$ indicate the source- and the target-domain datasets, respectively, where $N_\S$ is the number of source spectrograms and $N_\T$ the number of target spectrograms.
In the following, we assume that $\X^\S \cap \X^\T = \emptyset$.
Furthermore, we use $n$, $m$, $k$ as subscripts to index the tensors $\X^\S$ and $\X^\T$ in the 1\textsuperscript{st}, 2\textsuperscript{nd} and 3\textsuperscript{rd} dimension, respectively.
Finally, let each domain be characterized by a different distribution, i.e., let $\x_l^\S\sim\mathcal{X}^\S$ for $l=1,...,N_{\S}$ and $\x_\ell^\T\sim\mathcal{X}^\T$ for $\ell=1,...,N_{\T}$, where $\mathcal{X}^\S$ and $\mathcal{X}^\T$ are the source and target data distributions.

The proposed adaptation procedure comprises three steps.
First, we compute $\mu^\S_k$ and $\sigma^\S_k$ from $\X^\S$ as in:
\begin{equation}
    \mu^\S_k = \frac{1}{{N_\S}M}\sum_{n=1}^{N_\S}\sum_{m=1}^M \X^\S_{nmk}
    \label{eq:meanS}
\end{equation}
\begin{equation}
    \sigma^\S_k = \sqrt{\frac{1}{{N_\S}M-1}\sum_{n=1}^{N_\S}\sum_{m=1}^M \left(\X^\S_{nmk} - \mu^\S_k \right)^2}
    \label{eq:stdS}
\end{equation}
for $k=1,...,K$.
Intuitively, the values of $\mu_k^{\S}$ and $\sigma_k^{\S}$ are computed as the sample mean and standard deviation of the vector obtained by concatenating every $k$-th row of every spectrogram of $\X^{\S}$. 
Similarly, we compute $\mu_k^{\T}$ and $\sigma_k^{\T}$ for the target-domain dataset, i.e.,
\begin{equation}
    \mu^\T_k = \frac{1}{{N_\T}M}\sum_{n=1}^{N_\T}\sum_{m=1}^M \X^\T_{nmk}
    \label{eq:meanT}
\end{equation}
\begin{equation}
    \sigma^\T_k = \sqrt{\frac{1}{{N_\T}M-1}\sum_{n=1}^{N_\T}\sum_{m=1}^M \left(\X^\T_{nmk} - \mu^\T_k \right)^2}
    \label{eq:stdT}
\end{equation}
for $k=1,...,K$.

From this, we then standardize $\X^\T$ by setting
\begin{equation}
    \Z^\T_{nmk} = \frac{\left(\X^\T_{nmk} - \mu^\T_k\right)}{\sigma^\T_k}
    \label{eq:zscore}
\end{equation}
for $n=1,...,{N_\T}$, $m=1,...,M$ and $k=1,...,K$.

In the third step, the first- and second-order statistics of the source domain are finally used to transform the standardized target-domain data as follows:
\begin{equation}
    \overline{\X}^\T_{nmk} = \sigma^\S_k\Z^\T_{nmk} + \mu^\S_k
    \label{eq:invstd}
\end{equation}
for $n=1,...,{N_\T}$, $m=1,...,M$ and $k=1,...,K$.

At this point, ${\overline{\X}}^\T$ has been aligned to the source domain and shares the same band means and variances with $\X^\S$. Our hypothesis is that $\overline{\x}^\T \in \overline{\X}^\T$ would now be drawn from a distribution $\bar{\mathcal{X}}^\T$ which should be closer to $\mathcal{X}^\S$ than $\mathcal{X}^\T$, and thus that an ASC model trained on $\X^\S$ would achieve higher classification rates when evaluated on the aligned dataset $\overline{\X}^\T$ rather than on the non-adapted $\X^\T$.

Notably, the proposed method is model agnostic, does not involve any training and is completely unsupervised, i.e., it does not require target-domain labels at any given time.

\section{Evaluation}
\label{sec:eval}

\subsection{Training and Evaluation}
\label{ssec:models}

We evaluate the proposed adaptation procedure using two different ASC models. The first one, used also in \cite{mun2019domain}, is the baseline system of the DCASE 2018 Challenge \cite{mesaros2018multi} (denoted as ``DCASE model'' from now on). The second one, used in \cite{gharib2018unsupervised} and \cite{drossos2019unsupervised}, is the so-called ``Kaggle model''. The neural network architectures are implemented in PyTorch as described in Tables \ref{tab:dcasemodel} and \ref{tab:kagglemodel}.

To be consistent with the evaluation setup of the state of the art, the dataset used for the development and evaluation of the proposed approach is the one provided as the development dataset of Task 1-B of the DCASE 2018 Challenge \cite{mesaros2018multi}. The dataset contains 10-seconds long WAV files captured in six different large European cities using three different recording devices --- namely, devices A, B and C.
Each audio item is categorized by one of the ten scene labels:
\textit{airport}, \textit{bus}, \textit{metro}, \textit{metro station}, \textit{park}, \textit{public square}, \textit{shopping mall}, \textit{street pedestrian}, \textit{street traffic}, and \textit{tram}. 

Before being fed to the learning algorithms, the audio data is transformed into time-frequency features. Specifically, we extract log Mel-energies with different parameters depending on the model. For the DCASE model, we use 40 Mel-bands and a 40 ms Hamming window with 50\% overlap. For the Kaggle model, we use 64 Mel-bands and a 2048 samples ($\sim$~46 ms) Hamming window with 50\% overlap. We adopt the same training, validation and test folds as in \cite{gharib2018unsupervised}. In particular, the training set consists of 5510 audio clips (only Device A) and the validation set of 612 (only Device A). Furthermore, the test dataset contains 2878 clips recorded by different devices, namely 2518 files from Device A, 180 from Device B, and 180 from Device C.
Note that training data from Devices B and C is disregarded.

We optimize the DCASE model for 200 epochs using Adam.
The learning rate and the batch size were set to 10\textsuperscript{\textminus 4} and 16, respectively.
For the Kaggle model, to foster evaluation consistency and to show that our method can be effectively decoupled from the model training, we utilize the pre-trained weights made available online\footnote{\url{https://doi.org/10.5281/zenodo.1401995}} by the authors of \cite{gharib2018unsupervised}.

To be able to compare the results with prior works, we adopt an evaluation setup similar to the one of the DCASE 2018 Challenge, where the target-domain performance is assessed by averaging the accuracy obtained on Devices B and C. Hence, we evaluate our models on $\overbar\X^{\mathrm{(B,C)}}_{\mathrm{test}}$ and $\X^{\mathrm{(B,C)}}_{\mathrm{test}}$, i.e., the adapted and non-adapted test fold of Devices B and C combined. 
Moreover, to investigate the effect of the proposed adaptation method when applied to source-domain data, we evaluate our models using both $\overbar\X^{\mathrm{(A)}}_{\mathrm{test}}$ and $\X^{\mathrm{(A)}}_{\mathrm{test}}$, i.e., the adapted and non-adapted test fold of Device A, respectively. 

\subsection{Device-Dependent Adaptation}
\label{ssec:dda}

If we assume to have access to the knowledge of which device captured each acoustic scene in the target-domain dataset, i.e., if we assume that test samples are annotated with \textit{device labels}, we can consider the target devices separately during the adaptation phase. This means that Devices B and C are regarded as target domains in their own right and aligned independently from one another using the statistics of the source-domain training dataset (Device A), i.e.,
\begin{equation}
    \X^\S := \X^{\mathrm{(A)}}_\mathrm{train} \qquad \X^\T_\mathrm{B} := \X^{\mathrm{(B)}}_\mathrm{test} \qquad \X^\T_\mathrm{C} := \X^{\mathrm{(C)}}_\mathrm{test}
    \label{eq:defDDA}
\end{equation}
where $\X^{\mathrm{(A)}}_\mathrm{train}$ represents the training fold composed of $N_\S=\textrm{5510}$ samples from Device A, while $\X^{\mathrm{(B)}}_\mathrm{test}$ and $\X^{\mathrm{(C)}}_\mathrm{test}$ denote the test folds from Devices B and C, each consisting of $N_\T=\textrm{180}$ spectrograms.

Eventually, the two adapted target-domain datasets are concatenated to form the test dataset
\begin{equation}
    \overline{\X}_\mathrm{test}^{\mathrm{(B,C)}} := \left[\ \overline{\X}^\T_\mathrm{B}\,,\,\overline{\X}^\T_\mathrm{C}\ \right]\in\mathbb{R}^{2{N_\T} \times M \times K}
    \label{eq:testsetDDA}
\end{equation}
and the model evaluation can proceed as usual.
In the following, we refer to this approach as ``Device-Dependent Adaptation'' (DDA).

\subsection{Device-Independent Adaptation}
\label{ssec:dia}

To account for those cases in which the target-domain device labels are not available, we investigate the performance of the proposed adaptation method when one would consider a single domain comprising the features from both Device B and C. This means that the target domain consists of the concatenated dataset
\begin{equation}
    \X^\T_\mathrm{B|C} := \left[\ \X^{\mathrm{(B)}}_\mathrm{test}\,,\,\X^{\mathrm{(C)}}_\mathrm{test}\ \right]\in\mathbb{R}^{N'_\T \times M \times K}
    \label{eq:Xcat}
\end{equation}
where $N'_\T=360$ is the number of spectrograms in the target domain.
Then, having applied the proposed method to $\X^\T_\mathrm{B|C}$, we can define the adapted target-domain test dataset as
\begin{equation}
    \overbar\X_\mathrm{test}^{\mathrm{(B,C)}} := \overbar{\X}^\T_\mathrm{B|C}
    \label{eq:testsetDIA}
\end{equation}
In the following, we will refer to this approach as ``Device-Independent Adaptation'' (DIA).

\begin{table}[t]
    \caption{DCASE Model}
    \label{tab:dcasemodel}
    \centering
    \begin{tabular}{c}
        Input\\
        \scriptsize(40 Mel-bands)\\\toprule
        7$\times$7--Conv2D--32--BatchNormalization--ReLU\\
        5$\times$5--MaxPooling2D\\
        Dropout(.3)\\
        7$\times$7--Conv2D--64--BatchNormalization--ReLU\\
        4$\times$100--MaxPooling2D\\
        Dropout(.3)\\
        \midrule
        Dense--100--ReLU--Dropout(.3)\\\bottomrule
        Output--10--Softmax\\
    \end{tabular}
\end{table}
\begin{table}[t]
    \caption{Kaggle Model}
    \label{tab:kagglemodel}
    \centering
    \begin{tabular}{c}
        Input\\
        \scriptsize(64 Mel-bands)\\\toprule
        11$\times$11--Conv2D--48--stride(2,3)--padding(5)--ReLU\\
        3$\times$3--MaxPooling2D--stride(1,2)\\
        BatchNormalization\\
        5$\times$5--Conv2D--128--stride(2,3)--padding(2)--ReLU\\
        3$\times$3--MaxPooling2D--stride(2)\\
        BatchNormalization\\
        3$\times$3--Conv2D--192--stride(1)--padding(1)--ReLU\\
        3$\times$3--Conv2D--192--stride(1)--padding(1)--ReLU\\
        3$\times$3--Conv2D--128--stride(1)--ReLU\\
        3$\times$3--MaxPooling2D--stride(1,2)\\
        BatchNormalization\\\midrule
        Dense--256--ReLU--Dropout(.25)\\
        Dense--256--ReLU--Dropout(.25)\\\bottomrule
        Output--10--Softmax\\
    \end{tabular}
\end{table}

\begin{table*}[t]
    \caption{Classification accuracy on adapted and non-adapted source-domain (Device A) and target-domain test data (Devices B,\,C) obtained by both the DCASE and the Kaggle model. Note that two of the three variants of the method presented in \cite{mun2019domain} are not directly comparable with our approach and therefore they appear here in brackets (second and third row of the table).}
    \label{tab:results}
    \centering
    \begin{tabular*}{\linewidth}{@{\extracolsep{\fill}}lcccccccc}
    \toprule
    & \multicolumn{4}{c}{\textbf{DCASE Model}} & \multicolumn{4}{c}{\textbf{Kaggle Model}}\\
    \midrule
    & \multicolumn{2}{c}{\textbf{Non adapted}} & \multicolumn{2}{c}{\textbf{Adapted}} & \multicolumn{2}{c}{\textbf{Non adapted}} & \multicolumn{2}{c}{\textbf{Adapted}}\\
    & Device A & Devices B,C & Device A & Devices B,C & Device A & Devices B,C & Device A & Devices B,C\\
    \midrule
    \cite{mun2019domain} $\Delta\mu_2$\,derived\:from\:Device\:A  & --- & --- & 0.58 & 0.47 \\
    \cite{mun2019domain} $\Delta\mu_2$\,derived\:from\:Devices\:B,\,C  & --- & --- & (0.58) & (0.51) \\
    \cite{mun2019domain} Universal\;domain & --- & --- & (0.58) & (0.50) \\
    \cite{gharib2018unsupervised} & \multicolumn{4}{c}{} & 0.65 & 0.20 & 0.65 & 0.32 \\ 
    \cite{drossos2019unsupervised} & \multicolumn{4}{c}{} & 0.65 & 0.21 & 0.64 & 0.45 \\
    \midrule
    \textbf{Proposed\;method:} & \multicolumn{1}{c}{}\\
    \midrule
    DIA (Device-Independent\,Adaptation) & & & & 0.48 & & & & 0.45 \\
    DDA (Device-Dependent\:Adaptation) & \textbf{0.66} & 0.22 & 0.64 & \textbf{0.53} & 0.65 & 0.20 & \textbf{0.66} & \textbf{0.51} \\
    \bottomrule
    \end{tabular*}
\end{table*}

\subsection{Influence of the Number of Target-Domain Test Samples}
\label{ssec:number}

The proposed DDA and DIA methods have so far implied that $N_\T$ and $N'_\T$ target-domain samples would be available during the adaptation phase. 
However, many real-life applications cannot rely on such appropriately sized datasets.
In these cases, sample statistics are likely to be unreliable in describing device-specific characteristics, especially if the sample size is small. 
To assess how much the DA capabilities of the proposed method are affected by the amount of available data, we simulate the scenario in which only a limited number of target-domain samples are available. For the sake of brevity, we limit ourselves to the evaluation of the Kaggle model.

To this end, random permutations of the target-domain datasets $\X^\T_\mathrm{B}$, $\X^\T_\mathrm{C}$ and $\X^\T_\mathrm{B|C}$ are partitioned into segments of $L$ samples.
Specifically, $L$ is defined so that it takes values in the set of divisors of the number of spectrograms in the respective target-domain dataset, namely $N_\T$ for DDA and $N'_\T$ for DIA.
Each segment is then adapted independently of the others using the statistics of $\X^\S$ and subsequently concatenated to form $\overbar\X^\T_\mathrm{B}$, $\overbar\X^\T_\mathrm{C}$ and $\overbar\X^\T_\mathrm{B|C}$.
Finally, the test sets are obtained according to (\ref{eq:testsetDDA}) and (\ref{eq:testsetDIA}) and the evaluation of the DDA and DIA methods can proceed as described in the previous sections.
To reduce the influence of the random indexing involved in the segmentation process, we perform said procedure on 50 different permutations of each target dataset. Systems performance is then assessed by means of the average classification accuracy as a function of $L$.
Note, however, that each class has an equal probability of being represented within a segment of $L$ samples due to the preliminary random permutations. It is thus likely that multiple different acoustic scenes are contributing to the computation of the sample statistics, yielding a more robust estimate.

\section{Results and Discussion}
\label{sec:results}

Table \ref{tab:results} reports the results of the proposed method against the ones of the unsupervised methods in \cite{gharib2018unsupervised}, \cite{drossos2019unsupervised} and \cite{mun2019domain}. Note that, of the three variants presented in \cite{mun2019domain}, only the one that considers Device A as the reference domain used to train the FHVAE and to compute the channel conversion parameter $\Delta\mu_2$ is directly comparable with our method. Indeed, this is the only one not relying on external or target-domain data during the pre-training phase. 
For completeness, however, the results of the variants which do include such data are reported in brackets.

As can be seen from Table \ref{tab:results}, our DDA method provides a classification accuracy of 53\% (DCASE model) and 51\% (Kaggle model) when evaluated on the test fold of Devices B and C.
This corresponds to an increase of approximately 6\% in target-domain accuracy compared to \cite{mun2019domain} (47\%) and \cite{drossos2019unsupervised} (45\%).
For what concerns DIA, instead, we can notice that the performance matches the ones of much more complex systems based on FHVAE \cite{mun2019domain} and WGAN \cite{drossos2019unsupervised}. In particular, the accuracy obtained by DIA is 48\% (DCASE model) and 45\% (Kaggle model) against the 47\% of \cite{mun2019domain} and 45\% of \cite{drossos2019unsupervised}.
Moreover, when evaluated on Device A, the non-adapted procedure yields an accuracy of 66\% (DCASE model) and 65\% (Kaggle model), while the adapted one yields 64\% and 66\%, i.e., \textminus2\% and +1\%, respectively.
This slight mismatch is probably due to the sample statistics of training and test data being different and appears to be related to the number of Mel-bands chosen for the feature representations.

In view of the results, it seems that to apply such a band-wise preprocessing procedure across the datasets is beneficial when dealing with mismatched recording devices. 
A possible interpretation is that, while classic standardization approaches aim at balancing the weights of the features that describe each sample, our method is focusing more on device-specific characteristics which are constant throughout the dataset, rather than on the acoustic content of individual scenes. The result, in practice, is that of an equalization of the frequency response of the recording devices across domains.
This would also explain the lower classification accuracy when Devices B and C are combined into a single target-domain dataset before the adaptation (DIA), rather then preprocessed independently (DDA).
In the DIA case, indeed, the statistics removed by the standardization are only an average of the device-specific ones and therefore both channels would maintain a larger part of their characteristic response.

In a sense, the effect described here is somewhat reminiscent of the spectral correction presented in \cite{komider2019}, where the author finds as many coefficients as frequency bands from a reference device (Device A) which are then used to weight the time-frequency representations of audio data from the other devices. Notably, however, our method does not require pair-wise matchings between audio files recorded simultaneously.

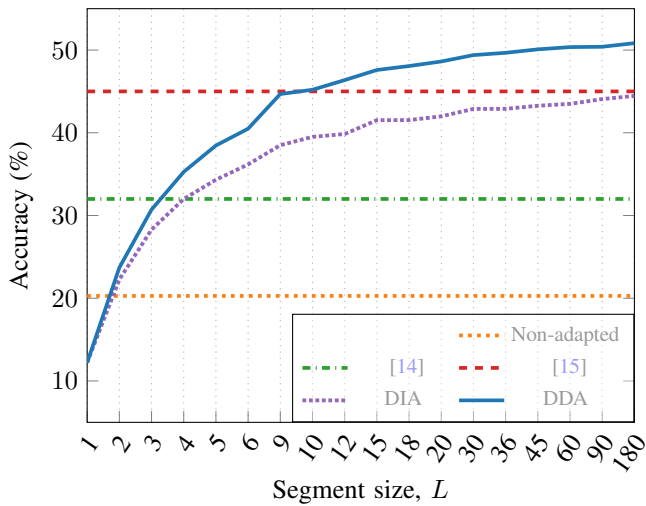
\begin{figure}[t]
    \centering
    \pgfplotstableread{abs/DDA.txt}\DDA%
\pgfplotstableread{abs/DIA.txt}\DIA%
\pgfplotstableread{abs/non_adapted.txt}\nonadapted%
\pgfplotstableread{abs/audasc.txt}\audasc%
\pgfplotstableread{abs/undaw.txt}\undaw%

\definecolor{c0}{HTML}{1F77B4}
\definecolor{c1}{HTML}{FF7F0E}
\definecolor{c2}{HTML}{2CA02C}
\definecolor{c3}{HTML}{D62728}
\definecolor{c4}{HTML}{9467BD}

\begin{tikzpicture}
    
   	\begin{axis}[enlargelimits=false,
    	legend columns=2,
        legend style={at={(1,0)},anchor=south east, font=\footnotesize, minimum height=0.43cm, minimum width=1.433cm},
        ymin=5, ymax=55,
        ytick={10,20,30,40,50},
        symbolic x coords={1,2,3,4,5,6,9,10,12,15,18,20,30,36,45,60,90,180},
        xtick=data,
	  	width=\columnwidth,
        height=0.8\columnwidth,
	  	xlabel={Segment size, $L$},
	    ylabel={Accuracy (\%)},
	    x label style={at={(axis description cs:.5,-.22)},anchor=south},
	    x tick label style={rotate=55,anchor=east, yshift=-5pt},
	    xtick pos=left,
	    ymajorgrids=false,
	    xmajorgrids=true,
        grid style={dotted, line width=.2pt, draw=gray!70},
	    legend style={fill=white, fill opacity=0.4, draw opacity=1, text opacity=1},
	    ]
	
	\addlegendimage{empty legend}\addlegendentry{}
    \addplot[color=c1, dotted, line width=1.4pt]table[x=0,y=1] 
    from \nonadapted;\addlegendentry{Non-adapted};
    \addplot[color=c2, dashdotted, line width=1.4pt]table[x=0,y=1] 
    from \audasc;\addlegendentry{\cite{gharib2018unsupervised}};
    \addplot[color=c3, dashed,  line width=1.4pt]table[x=0,y=1] 
    from \undaw;\addlegendentry{\cite{drossos2019unsupervised}};
    \addplot[color=c4, densely dotted, line width=1.45pt]table[x=0,y=1]
    from \DIA;\addlegendentry{DIA}
    \addplot[color=c0, solid, line width=1.4pt]table[x=0,y=1]
    from \DDA;\addlegendentry{DDA}
    \end{axis} 
\end{tikzpicture}
    \vspace{-12pt}
    \caption{Average accuracy of the Kaggle model on $\overbar\X^{\mathrm{(B,C)}}_\mathrm{test}$ as a function of the number of samples in each test segment. The figure depicts the results of both DDA (blue solid curve) and DIA (purple dotted curve) plotted against the performance of the Kaggle model tested on non-adapted data (orange dotted~line), \cite{gharib2018unsupervised} (green dash-dotted line) and \cite{drossos2019unsupervised} (red~dashed line).}
    \label{fig:abs}
    \vspace{-0.1em}
\end{figure}

As mentioned in \cite{mun2019domain}, a desired quality of unsupervised adaptation strategies is not to rely on a large-sized target-domain dataset. In the following, we show that our method is capable of providing satisfactory results even when just a few target-domain samples are available.
By looking at \figurename\,\ref{fig:abs}, we can readily observe that, as expected, classification accuracy grows monotonically with $L$ and the maximum is achieved for a segment encompassing all the possible acoustic scenes, i.e., $L=N_\T$ (DDA) and $L=N'_\T$ (DIA). 
On the one hand, $L=1$ (corresponding to trying to align every Mel-spectrogram independently of the others) is worse than applying no adaptation at all.
On the other hand, the proposed DDA method is already capable of outperforming \cite{drossos2019unsupervised} using segments of $L\ge10$ samples (i.e., just over one and a half minutes of audio). 
This is quite remarkable as it suggests that considerable DA can be achieved without the burden of gathering an abundance of target-domain samples, making it feasible for a user to collect the data needed to adapt their own device.
However, a more thorough study on the effect of under-representation of certain classes in the test segments is left for future work.

\section{Conclusions and Future Work}
\label{sec:conclusions}

We proposed an effective approach to unsupervised domain adaptation for acoustic scene classification. Our method, despite its simplicity, is able to outperform the unsupervised methods found in the literature while requiring just over ten test samples to provide state-of-the-art results. Moreover, we showed that our approach is competitive in terms of target-domain classification accuracy even without being given the knowledge of which target device captured which acoustic scene.
The adaptation procedure is computationally efficient, model agnostic and does not involve any training. Therefore, our method can be readily applied to any previously optimized ASC model without the need of further adjustments.
For future work, we plan to evaluate the proposed method for other audio classification tasks, such as speech and music recognition.

\bibliographystyle{IEEEtran}
\bibliography{IEEEabrv,refs}

\begin{thebibliography}{10}
\providecommand{\url}[1]{#1}
\csname url@samestyle\endcsname
\providecommand{\newblock}{\relax}
\providecommand{\bibinfo}[2]{#2}
\providecommand{\BIBentrySTDinterwordspacing}{\spaceskip=0pt\relax}
\providecommand{\BIBentryALTinterwordstretchfactor}{4}
\providecommand{\BIBentryALTinterwordspacing}{\spaceskip=\fontdimen2\font plus
\BIBentryALTinterwordstretchfactor\fontdimen3\font minus
  \fontdimen4\font\relax}
\providecommand{\BIBforeignlanguage}[2]{{%
\expandafter\ifx\csname l@#1\endcsname\relax
\typeout{** WARNING: IEEEtran.bst: No hyphenation pattern has been}%
\typeout{** loaded for the language `#1'. Using the pattern for}%
\typeout{** the default language instead.}%
\else
\language=\csname l@#1\endcsname
\fi
#2}}
\providecommand{\BIBdecl}{\relax}
\BIBdecl

\bibitem{mesaros2018detection}
A.~{Mesaros}, T.~{Heittola}, E.~{Benetos}, P.~{Foster}, M.~{Lagrange},
  T.~{Virtanen}, and M.~D. {Plumbley}, ``Detection and classification of
  acoustic scenes and events: Outcome of the {DCASE} 2016 challenge,''
  \emph{IEEE/ACM Trans. Audio, Speech, Language Process.}, vol.~26, no.~2, pp.
  379--393, 2018.

\bibitem{mesaros2018acoustic}
A.~{Mesaros}, T.~{Heittola}, and T.~{Virtanen}, ``Acoustic scene
  classification: An overview of {DCASE} 2017 challenge entries,'' in
  \emph{Proc. of the International Workshop on Acoustic Signal Enhancement
  (IWAENC)}, 2018, pp. 411--415.

\bibitem{sakashita2018}
Y.~Sakashita and M.~Aono, ``Acoustic scene classification by ensemble of
  spectrograms based on adaptive temporal divisions,'' Detection and
  Classification of Acoustic Scenes and Events 2018 Challenge (DCASE2018),
  Tech. Rep., 2018.

\bibitem{koutini2019receptive}
K.~{Koutini}, H.~{Eghbal-zadeh}, M.~{Dorfer}, and G.~{Widmer}, ``The receptive
  field as a regularizer in deep convolutional neural networks for acoustic
  scene classification,'' in \emph{Proc. of the European Signal Processing
  Conference (EUSIPCO)}, 2019, pp. 1--5.

\bibitem{ben2010theory}
S.~Ben-David, J.~Blitzer, K.~Crammer, A.~Kulesza, F.~Pereira, and J.~W.
  Vaughan, ``A theory of learning from different domains,'' \emph{Machine
  Learning}, vol.~79, no.~1, pp. 151--175, 2010.

\bibitem{daume2006domain}
H.~Daum{\'e}~III and D.~Marcu, ``Domain adaptation for statistical
  classifiers,'' \emph{J. Artif. Intell. Res.}, vol.~26, pp. 101--126, 2006.

\bibitem{blitzer2006domain}
J.~Blitzer, R.~McDonald, and F.~Pereira, ``Domain adaptation with structural
  correspondence learning,'' in \emph{Proc. of the Conference on Empirical
  Methods in Natural Language Processing}, 2006, p. 120–128.

\bibitem{blitzer2007biographies}
J.~Blitzer, M.~Dredze, and F.~Pereira, ``Biographies, {Bollywood}, boom-boxes
  and blenders: Domain adaptation for sentiment classification,'' in
  \emph{Proc. of the annual meeting of the association of computational
  linguistics}, 2007, pp. 440--447.

\bibitem{glorot2011domain}
X.~Glorot, A.~Bordes, and Y.~Bengio, ``Domain adaptation for large-scale
  sentiment classification: A deep learning approach,'' in \emph{Proc. of the
  International Conference on Machine Learning (ICML)}, 2011, pp. 513--520.

\bibitem{saenko2010adapting}
K.~Saenko, B.~Kulis, M.~Fritz, and T.~Darrell, ``Adapting visual category
  models to new domains,'' in \emph{Proc. of the European Conference on
  Computer Vision (ECCV)}, 2010, pp. 213--226.

\bibitem{hoffman2017cycada}
J.~Hoffman, E.~Tzeng, T.~Park, J.-Y. Zhu, P.~Isola, K.~Saenko, A.~Efros, and
  T.~Darrell, ``{C}y{CADA}: Cycle-consistent adversarial domain adaptation,''
  in \emph{Proc. of the International Conference on Machine Learning (ICML)},
  2018, pp. 1989--1998.

\bibitem{mesaros2018multi}
A.~Mesaros, T.~Heittola, and T.~Virtanen, ``A multi-device dataset for urban
  acoustic scene classification,'' in \emph{Proc. of the Detection and
  Classification of Acoustic Scenes and Events 2018 Workshop (DCASE2018)},
  2018, pp. 9--13.

\bibitem{mesaros2019acoustic}
------, ``Acoustic scene classification in {DCASE} 2019 challenge: Closed and
  open set classification and data mismatch setups,'' in \emph{Proc. of the
  Detection and Classification of Acoustic Scenes and Events 2019 Workshop
  (DCASE2019)}, 2019, pp. 164--168.

\bibitem{gharib2018unsupervised}
S.~Gharib, K.~Drossos, E.~\c{C}akir, D.~Serdyuk, and T.~Virtanen,
  ``Unsupervised adversarial domain adaptation for acoustic scene
  classification,'' in \emph{Proc. of the Detection and Classification of
  Acoustic Scenes and Events 2018 Workshop (DCASE2018)}, 2018, pp. 138--142.

\bibitem{drossos2019unsupervised}
K.~{Drossos}, P.~{Magron}, and T.~{Virtanen}, ``Unsupervised adversarial domain
  adaptation based on the {Wasserstein} distance for acoustic scene
  classification,'' in \emph{Proc. of the Workshop on Applications of Signal
  Processing to Audio and Acoustics (WASPAA)}, 2019, pp. 259--263.

\bibitem{mun2019domain}
S.~Mun and S.~Shon, ``Domain mismatch robust acoustic scene classification
  using channel information conversion,'' in \emph{Proc. of the International
  Conference on Acoustics, Speech and Signal Processing (ICASSP)}, 2019, pp.
  845--849.

\bibitem{tzeng2017adversarial}
E.~Tzeng, J.~Hoffman, K.~Saenko, and T.~Darrell, ``Adversarial discriminative
  domain adaptation,'' in \emph{Proc. of the Conference on Computer Vision and
  Pattern Recognition (CVPR)}, 2017, pp. 7167--7176.

\bibitem{arjovsky2017wasserstein}
M.~Arjovsky, S.~Chintala, and L.~Bottou, ``Wasserstein generative adversarial
  networks,'' in \emph{Proc. of the International Conference on Machine
  Learning (ICML)}, 2017, pp. 214--223.

\bibitem{komider2019}
M.~Kośmider, ``Calibrating neural networks for secondary recording devices,''
  Detection and Classification of Acoustic Scenes and Events 2019 Challenge
  (DCASE2019), Tech. Rep., 2019.

\end{thebibliography}

\end{document}